# Experimental Framework for Generating Reliable Ground Truth for Laryngeal Spatial Segmentation Tasks

**Hamzeh Ghasemzadeh** [a,b,c], **David S. Ford** [d], **Maria E. Powell** [e], **Dimitar D. Deliyski** [c]

[a] Center for Laryngeal Surgery and Voice Rehabilitation, Massachusetts General Hospital, Boston, MA, USA
[b] Department of Surgery, Harvard Medical School, Boston, MA, USA
[c] Department of Communicative Sciences and Disorders, Michigan State University, East Lansing, MI, USA
[d] Department of Speech-Language Pathology, John G. Rangos, Sr. School of Health Sciences, Duquesne University, Pittsburgh, PA, USA
[e] Department of Otolaryngology - Head & Neck Surgery, Vanderbilt University Medical Center, Nashville, TN, USA

Running Title: Framework for Spatial Segmentation Ground Truth

Email Address of the Authors:
Hamzeh Ghasemzadeh, email: hghasemzadeh@mgh.harvard.edu

Corresponding Author:
Hamzeh Ghasemzadeh
Center for Laryngeal Surgery and Voice Rehabilitation
Massachusetts General Hospital
One Bowdoin Square, 11th Floor
Boston, MA 02114

ABSTRACT:
**Objective:** The validity of objective measures derived from high-speed videoendoscopy (HSV) depends, among other factors, on the validity of spatial segmentation. Evaluation of the validity of spatial segmentation requires the existence of reliable ground truths. This study presents a framework for creating reliable ground truth with sub-pixel resolution and then evaluates its performance.

**Method**: The proposed framework is a three-stage process. First, three laryngeal imaging experts performed the spatial segmentation task. Second, regions with high discrepancies between experts were determined and then overlaid onto the segmentation outcomes of each expert. The marked HSV frames from each expert were randomly assigned to the two remaining experts, and they were tasked to make proper adjustments and modifications to the initial segmentation within disparity regions. Third, the outcomes of this reconciliation phase were analyzed again and regions with continued high discrepancies were identified and adjusted based on the consensus among the three experts. This three-stage framework was tested using a custom graphical user interface that allowed precise piece-wise linear segmentation of the vocal fold edges. Inter-rate reliability of segmentation was evaluated using 12 HSV recordings. 10% of the frames from each HSV file were randomly selected to assess the intra-rater reliability.

**Result and conclusion**: The reliability of spatial segmentation progressively improved as it went through the three stages of the framework. The proposed framework generated highly reliable and valid ground truths for evaluating the validity of automated spatial segmentation methods.

INTRODUCTION

Laryngeal images provide a direct method for studying and assessing laryngeal physiology and pathophysiology and are an important part of instrumental assessment of voice [1]. High-speed videoendoscopy (HSV) is the imaging modality that captures images at sampling rates higher than 4,000 frames per second (fps) [2] and it offers images with high spatial and high temporal resolutions. HSV is ideal for studying and evaluating the intra-cycle variation of vocal fold vibration, transient phenomena (e.g., voice onset, voice offset, voice breaks, glides, etc.), and non-periodic phonation (e.g., voices with severe dysphonia) [3]. Additionally, the combination of this wealth of temporal and spatial information with spatial calibrated measurements [4], [5] and three-dimensional reconstruction [6], [7] capabilities can open up many new possibilities for precision medicine and modeling applications. However, the existence of a reliable and robust automated processing pipeline is the prerequisite for achieving the full potential of HSV.

Motion compensation [8], temporal segmentation [9], spatial calibration [10], and spatial segmentation [11] are among the most common components of a HSV processing pipeline. The purpose of motion compensation is to account for the movement of the endoscope during the recording by bringing the region of interest to a fixed window [8]. Temporal segmentation detects and locates the time points that the target phonatory gestures (e.g., onset, phonation, voice break, offset, etc.) were happening in the recording [9]. The size of laryngeal tissue (e.g., vocal folds or a lesion) in laryngeal images depends on their distances from the endoscope [7] and the imaging angle [12]. Spatial calibration is the process that accounts for those confounding factors. Last but not least, spatial segmentation is the component that detects the region of interest (often the vocal folds) in HSV recordings. A comprehensive review of spatial segmentation methods can be found in [11].

Spatial segmentation is the most ubiquitous component of any HSV processing pipeline, and its accuracy is crucial to the accuracy and validity of any measure computed from HSV. Hence, rigorous evaluation of spatial segmentation is necessary, which requires the existence of valid and reliable ground truths. Spatial segmentation ground truth comes from manual segmentations by experts, which is a subjective task. Several aspects need to be considered for the proper evaluation of a spatial segmentation method. First, multiple experts need to participate in the segmentation task to ensure the generation of reliable ground truths. Second, similar to any other measure that relies on human judgment, the adopted methodology for generating the ground truth should include the evaluation of both inter- and intra-rater reliability. Third, the ground truth dataset should include recordings from patients and controls, as well as from males and females to account for possible effects of pathology and sex. Fourth, the resolution of the ground truth should be appropriate for the target application. This means that while segmentation with pixel resolution may be fine for HSV measures that are computed from the glottal area waveform, velocity measures [13], [14], [15], [16], [17] need appropriate ground truth with sub-pixel resolution. Fifth, data leakage should be prevented in machine learning-based spatial segmentation methods by keeping the ground truth samples in the test set completely separate from the validation and training set [18]. Nested cross-validation can achieve this very effectively [18].

Reviewing the literature on spatial segmentation shows a general lack of rigorous evaluation methodology. Several studies didn't evaluate the accuracy of their spatial segmentation at all [19], [20], [21]. Evaluation of other studies was limited to just four points on the vocal fold edges [22], or only the mid cross-section of the vocal fold [23], which could not reflect the true performance of these methods. The ground truth in some studies was not generated independently. For example, in [23], the segmentation outcome was overlaid on images, and experts were then asked to evaluate the accuracy of segmentation, which most likely can bias the evaluation. In [24] the initial segmentations were done automatically, and then experts were asked to adjust them. While this approach is acceptable for generating a large database for training deep neural networks, it might give a biased evaluation of the performance of spatial segmentation. Another finding was the absence of an evaluation of inter- and intra-rater reliability in many of them. For example, among the investigated spatial segmentation studies only three included some evaluation of inter-rater reliability [24], [25], [26]. The only study that evaluated intra-rate reliability was [26], which was the segmentation of the glottic angle and not the vocal fold edges or the glottis. The number of experts segmenting a frame was not mentioned in some of the studies [23], [27]; it was one in [24], [28], with the study [24] including post-hoc checking and modification by a second expert. The dataset for most studies was very small and didn't include adequate variabilities in terms of patients, controls, males, and females. For example, studies [23], [28] were based on one female control, while other studies only included patients or controls [27], [29], [30]. It is noteworthy that the study presented in [24] was an exception, and it included both patients and controls and males and females. Last but not least, the evaluation of spatial segmentation methods that offer sub-pixel resolution [10], [23], [27] needs reliable ground truth with a sub-pixel resolution, which currently is not available.

The present study addressed some of these identified gaps. The first aim of this study was to quantify possible differences between the manual segmentation of different experts and hence to highlight the necessity of participation of multiple raters during the manual segmentation process. The second aim of this study was to present a framework that would allow multiple experts to participate in the manual segmentation of laryngeal images in an iterative and multi-stage procedure. This framework incorporates the evaluation of inter- and intra-rater reliability and governs the generation of very reliable ground truth with sub-pixel resolution. The growing interest in studying vocal fold velocity [10], [14], [15], [16], [17], [31] highlights the importance of segmentation with sub-pixel resolution capabilities. The framework is implemented and evaluated with three experts (all obtained their advanced degrees with a focus on the analysis and interpretation of HSV) for generating reliable ground truth for vocal fold edges. Application of the proposed framework is not limited to segmentation of the vocal fold edges and can be used for evaluation of segmentation of glottal angle [15], [16], [17], [26], segmentation of vocal fold lesions [10], [32], laryngeal segmentation [33], or any other spatial segmentation task.

## MATERIALS AND METHODS
### The proposed framework

The proposed framework has three stages, and it is an iterative process that has been designed to create reliable grand truth with sub-pixel resolution for spatial segmentation tasks. A graphical user interface (GUI) with zooming capability was developed to assist segmentation at different stages of the process. First, three laryngeal imaging experts independently segmented vocal fold edges from HSV recordings. Experts were encouraged to use the capabilities of the GUI and move backward and forward between frames and to use the temporal information (i.e., movement of vocal fold edges) during the segmentation of each frame. The outcome of this phase gave us three different estimates of the grand truth. Second, the three initial estimates underwent a reconciliation stage. All initial estimates were processed, and the regions with discrepancies higher than a threshold value (T>0.5 pixels) were identified and marked by a rectangle on the segmentation outcomes of all three experts. Then, the marked frames from each expert were randomly divided between the other two experts, meaning 50% of frames segmented by expert1 were assigned to expert2, and the rest were assigned to expert3. The same process was repeated for frames segmented by the other two experts. At this stage, each expert was asked to review the regions marked as areas with high discrepancies and either adjust it or accept it. Experts only saw one of the three initial segmentations at this stage and were blinded to the other two initial segmentations. Third, the three reconciled estimates were overlaid on top of each other, and then the regions with discrepancies higher than a threshold value (T>0.5 pixels) were identified and marked by a rectangle. The result of this stage was presented to all three experts, and they were asked to look simultaneously at the three reconciled segmentation outcomes and arrive at a consensus on where the edges need to be within the discrepancy regions. Figure 1 represents a flow chart of the proposed framework.

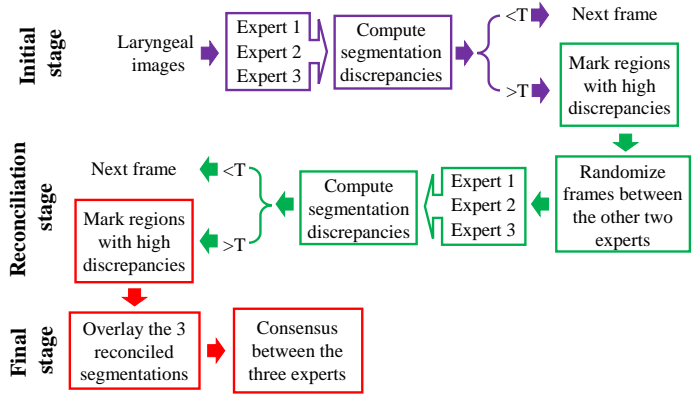

Figure 1. Flowchart of the proposed framework.

**Dataset and evaluation methodology**

Our developed GUI allowed the experts to use piecewise linear approximation and mark the edges of the vocal folds. Any complex contour can be segmented using this approach if enough anchor points are used. Figure 2(A) shows an example of segmentation. The GUI was equipped with zooming capabilities that allowed experts to see the pixelated image and assisted them in seeing where exactly they were placing each anchor. Also, the GUI allowed modifying the existing anchors and breaking a piece-wise linear segment into two segments by inserting a new anchor between any two existing anchors. Laryngeal images were taken from consecutive frames of HSV recordings. The experts were able to move between frames and exploit the temporal information to achieve a more precise segmentation outcome. The threshold for discrepancy region was selected as T=0.5 pixel, meaning discrepancy regions were scanning lines (the anterior-posterior direction) that one of the experts had their segmentation at least 0.5 pixels more lateral or medial than the other two experts. Such a region also needed to be at least two pixels long (anterior-posterior direction) to be marked as a discrepancy region. Figure 2(B) shows the segmentation outcome from three experts and a marked discrepancy region.

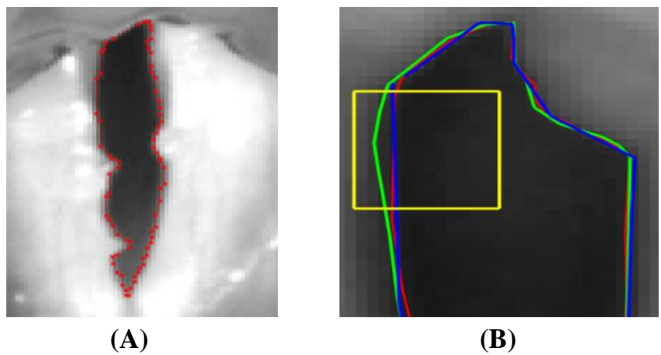

Figure 2. A: Piece-wise linear approximation of vocal edges with adequate anchor points can segment very complex contours. B: Segmentation outcome from three experts and a marked discrepancy region.

The method was evaluated using 12 HSV recordings from the BaGLS dataset [24] in a 2×2 design (male vs. female, and patient vs. control) with 3 samples per each group. HSV data were acquired using a monochrome camera at the temporal resolution of 4,000 frames per second (fps) and a rigid endoscope. Eleven of the recordings had a spatial resolution of 512×256 pixels, and the remaining one had a spatial resolution of 320×256 pixels. Two of the participants were in the age range of 10-20 years, five were in the age range of 20-30 years, one was in the age range of 40-50 years, two were in the age range of 50-60 years, and the remaining two were in the age range of 70-80 years. 100 consecutive frames from each recording were selected randomly for the segmentation task, among which 10 frames were selected randomly and were re-segmented using the same three-staged framework (10% redundancy) to allow computation of intra-rater reliability.

**Generation of the ground truth**

Experts segmented edges based on piece-wise linear approximation. The points on each line were quantized with a resolution of 0.1 pixels. Therefore, the edges had a resolution of 0.1 pixels in the medial-lateral and anterior-posterior directions. The medial-lateral coordinate of each edge of the vocal fold had three different estimates per scanning line (each coming from the segmentation of one of our experts). The average of the three estimates was used as the medial-lateral coordinate of the ground truth for that scanning line. This process was repeated separately for all scanning lines of each edge of the vocal folds.

**Evaluation criteria**

Intersection over union (IOU) is the most widely used metric for the evaluation of segmentation accuracy in image processing [34]. Figure 3(A) gives an example of how IOU is computed. Let the yellow solid circle and the green dashed circle denote two possible segmentations of the target region (glottis in our case). IOU is defined as the area of the intersection of the two regions (the red region) divided by the area of their union (the blue region). Equation 1 shows this. The IOU score is a number between 0 and 1, and a larger number indicates a higher agreement between the two possible segmentations. IOU can be used for the evaluation of inter- and intra-rater reliability. In the intra-rater case, we had two different segmentations from each expert and IOU evaluated the similarity between the original segmentation and the repeated segmentation. In the inter-rater case, there were three possible segmentations per frame, and we wanted to assess how similar they were to each other. Therefore, IOU was computed for different possible combinations of experts (i.e., Expert1-Expert2, Expert1-Expert3, and Expert2-Expert3) and then they were averaged.

$$IOU = \frac{Area\ (red)}{Area\ (blue)} \qquad (1)$$

Investigating the equation of IOU suggests that it may be sensitive to the area of the target region. Specifically, let's assume that an expert has a consistent average segmentation error of 1 pixel, meaning on average, they will draw the boundary of an object 1 pixel away from its actual location. Now if we ask the same expert to segment objects with different sizes, the smaller object will have a smaller IOU even though the segmentation error of the expert has been consistent. This inherent limitation of IOU could especially be problematic for the evaluation of segmentation of vocal fold edges, as the area of the glottis changes during phonation. Figure 3(B) shows an example scatter plot of the computed IOU for different areas of the glottis, which clearly shows this phenomenon.

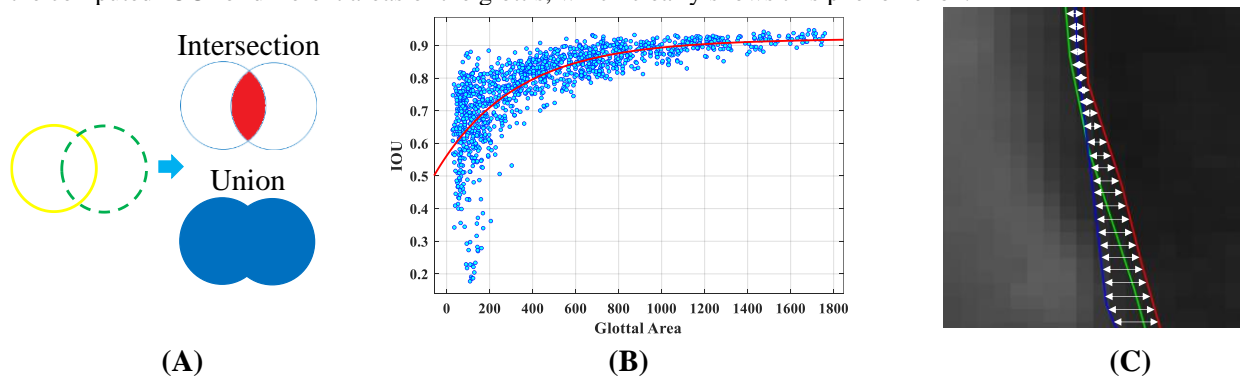

(A)　　　　　(B)　　　　　(C)

Figure 3. A: Definition and computation of Intersection Over Union (IOU). B: Association between IOU score and the area of the target region (glottis). C: Definition and computation of maximum edge variability (EV$_{max}$).

This study proposes edge variability as an alternative evaluation criterion; however, for the sake of completeness and given the ubiquitous presence of IOU in the literature, we will still report IOU in this paper. Edge variability quantifies lateral differences between different segmentations of the edges over different scanning lines (anterior-posterior sections) of the vocal folds, that is, how close or far different segmentations of the vocal folds were throughout the length of the glottis. The definition and computation of lateral differences between two possible segmentations (e.g., intra-rater and pair-wise inter-rater evaluation) is trivial. However, when there are multiple segmentations lateral differences need to be defined. Two different metrics of average maximum edge variability (EV$_{max}$) and average median edge variability (EV$_{median}$) were defined in this study. Average maximum edge variability (EV$_{max}$) was defined as the "maximum" lateral difference between segmentations of vocal fold edges among all experts averaged over all scanning lines of the vocal folds (the average of the lengths of all white arrows in Figure 3(C)). The value of average maximum edge variability could be dominated by the segmentation outcome of one of the experts if they consistently would have segmented the edge differently compared to other experts. Average median edge variability (EV$_{median}$) can prevent such instances, and it was defined as the median of lateral differences in segmentation of all possible pair-wise combinations of experts (Expert1-Expert2, Expert1-Expert3, and Expert2-Expert3) averaged over all scanning lines of the vocal folds. It must be noted that when there are two possible segmentations (e.g., intra-rater and pair-wise inter-rater evaluation) $EV_{max}$ and $EV_{median}$ will be identical, and hence only $EV_{max}$ was reported in this study.

Edge variability can assess inter-rater reliability, intra-rater reliability, and segmentation uncertainty. In the intra-rater case, we had two different segmentations from each expert and $EV_{max}$ evaluated the similarity between the original segmentation and the repeated segmentation. In the inter-rater case, there were three possible segmentations per frame, and we wanted to assess how similar they were to each other. Similar to the IOU case, we computed $EV_{max}$ for different possible combinations of experts (i.e., Expert1-Expert2, Expert1-Expert3, and Expert2-Expert3) and then averaged them. Uncertainty is a non-negative number that relates to the measure dispersion (i.e., variability) [35]. $EV_{max}$ and $EV_{median}$ computed from the segmentation of all experts quantify the variability of segmentation between different experts and can assess the uncertainty of the ground truth generated from the segmentation of experts. More specifically, a lower value of $EV_{max}$ or $EV_{median}$ means that the three experts were marking the edges more closely to each other, and hence the dispersion of the segmentation had been lower. This would translate into lower uncertainty of spatial segmentation, and hence higher confidence and reliability of the generated ground truth.

**Statistical analysis**

Our target measures (IOU, $EV_{max}$, $EV_{median}$) had non-gaussian distributions based on the Shapiro-Wilk test (p<0.00001); therefore, the Kruskal-Wallis test formally evaluated significant differences between different conditions. The effect size of significant differences was quantified using $\varepsilon^2$, and Tukey's honestly significant difference procedure was used for post-hoc analysis.

RESULTS

Two experiments were conducted in this study. Experiment 1 quantifies differences in the manual segmentation of different experts and demonstrates the performance of the proposed framework in the presence of such differences. Experiment 2 quantifies the reliability of the ground truth generated at different stages of the proposed framework.

**Experiment1: reliability of different experts**

First, the intra-rater reliability of different experts at different stages of the framework was calculated using the redundant 120 frames (10 frames per video recording). Figure 4 presents distributions of IOU and $EV_{max}$ for each expert after the initial and reconciliation stages. Please note that for intra-rater evaluation, two segmentations will be compared and hence $EV_{median}$ and $EV_{max}$ will be the same. Therefore, $EV_{median}$ was not reported in this analysis.

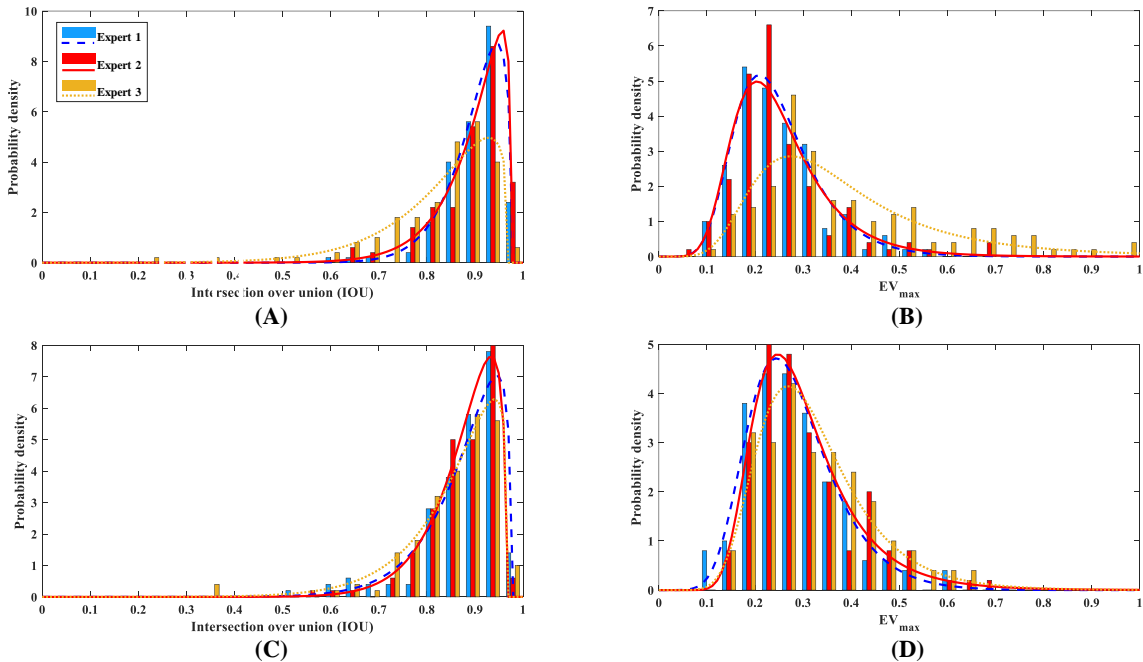

Figure 4. Intra-rater reliability of each expert. (A) IOU score after the initial segmentation, (B) $EV_{max}$ score after the initial segmentation, (C) IOU score after the reconciliation phase, (D) $EV_{max}$ score after the reconciliation phase.

Figures 4-A and 4-C suggest that the third expert had lower intra-rater reliability scores (lower IOU and higher $EV_{max}$) compared to the other two experts during the initial phase. However, based on Figures 4-B and 4-D segmentations of expert 3 became consistent with the segmentation of the other two experts following the reconciliation phase. Intra-rater reliability scores had non-gaussian distributions based on the Shapiro-Wilk test ($p<0.00001$); therefore, the Kruskal-Wallis test formally evaluated significant differences between the reliability scores of different experts. The independent variable was one of the reliability scores (initial IOU, reconciliation IOU, initial $EV_{max}$, and reconciliation $EV_{max}$), and the dependent variable was different experts. There was a significant difference in initial IOU across the experts ($\chi^2(2) = 41.5$, $p<0.00001$) with a moderate effect size ($\varepsilon^2=0.12$). Post-hoc analysis confirmed significantly lower IOU for expert3 compared to both expert1 and expert2. However, there was no significant difference in reconciliation IOU across the experts ($\chi^2(2) = 4.1$, $p=0.13$). Similarly, there was a significant difference in initial $EV_{max}$ across the experts ($\chi^2(2) = 61.2$, $p<0.00001$) with a relatively strong effect size ($\varepsilon^2=0.17$). Post-hoc analysis confirmed significantly higher $EV_{max}$ for expert3 compared to both expert1 and expert2. There was also a marginally significant difference in reconciliation $EV_{max}$ across the experts ($\chi^2(2) = 7.3$, $p=0.03$) with a weak effect size ($\varepsilon^2=0.02$). It must be noted that the difference became non-significant after accounting for multiple comparisons ($\alpha=0.05/4= 0.012$).

Second, the pair-wise inter-rater reliability between different experts at different stages of the framework was calculated. Figure 5 presents distributions of IOU and $EV_{max}$ between different pairs of experts after the initial and reconciliation stages. Please note that in pair-wise intra-rater evaluation, two segmentations will be compared at a time and hence, $EV_{median}$ and $EV_{max}$ will be the same. Therefore, $EV_{median}$ was not reported in this analysis.

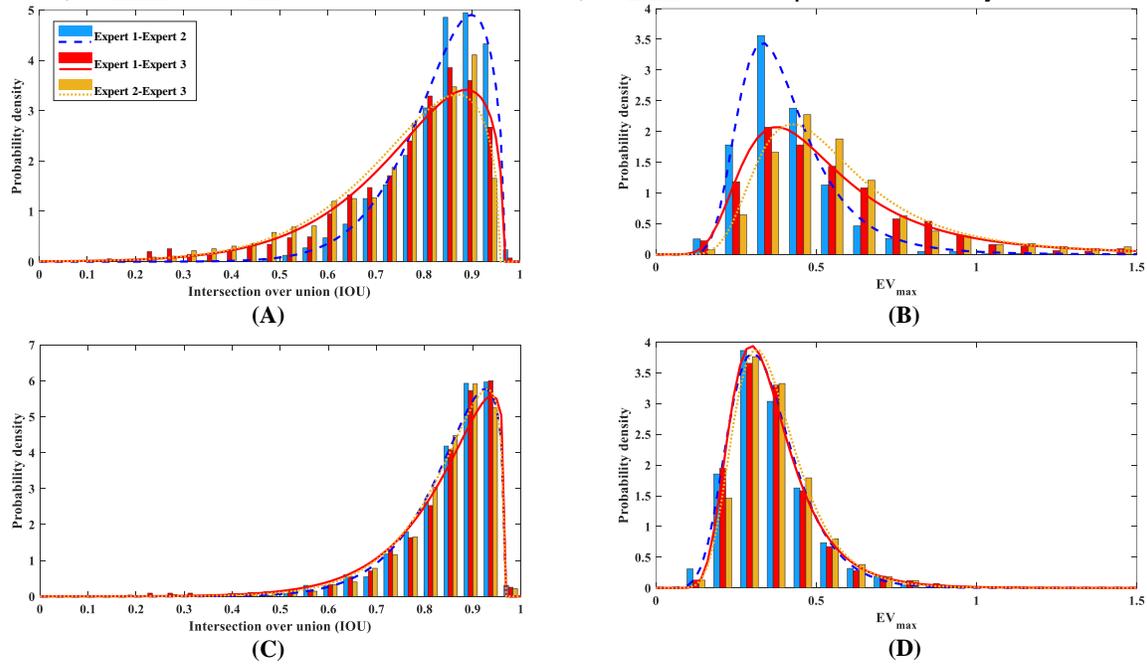

Figure 5. The pair-wise inter-rater reliability. (A) IOU score after the initial segmentation, (B) $EV_{max}$ score after the initial segmentation, (C) IOU score after the reconciliation phase, (D) $EV_{max}$ score after the reconciliation phase.

Figures 5-A and 5-C suggest that the third expert had lower pair-wise inter-rater reliability scores (lower IOU and higher $EV_{max}$) compared to the pair-wise inter-rater reliability scores of the other two experts during the initial phase. However, based on Figures 5-B and 5-D segmentations of expert 3 became consistent with the segmentation of the other two experts following the reconciliation phase. Inter-rater reliability scores had non-gaussian distributions based on the Shapiro-Wilk test ($p<0.00001$); therefore, the Kruskal-Wallis test formally evaluated significant differences between pair-wise reliability scores of different experts. The independent variable was one of the reliability scores (initial IOU, reconciliation IOU, initial $EV_{max}$, and reconciliation $EV_{max}$), and the dependent variable was different experts. There was a significant difference in initial IOU across the experts ($\chi^2(2) = 211.7$, $p<0.00001$) with a small effect size ($\varepsilon^2=0.05$). Post-hoc analysis confirmed significantly different scores across the three pair-wise evaluations, with expert 1-expert 2 showing the most distinct difference compared to the other two cases. lower IOU for expert3 compared to both expert1 and expert2. However, there was no significant difference in pair-wise reconciliation IOU across the experts ($\chi^2(2) = 3.8$, $p=0.15$). Similarly, there was a significant difference in initial $EV_{max}$ across the experts ($\chi^2(2) = 434.7$, $p<0.00001$) with a moderate effect size ($\varepsilon^2=0.11$). Post-hoc analysis

confirmed significantly different scores across the three pair-wise evaluations, with expert 1-expert 2 showing the most distinct difference compared to the other two cases. There was also a significant difference in reconciliation $EV_{max}$ across the pair-wise evaluations ($\chi^2(2) = 14.2$, $p=0.0008$) however with only a negligible effect size ($\varepsilon^2= 0.004$).

**Experiment2: reliability of the ground truth**

The uncertainty of the ground truth at different stages of the framework was quantified using $EV_{max}$ and $EV_{median}$ across the three experts. In that sense, a lower value of $EV_{max}$ or $EV_{median}$ indicates the three experts were marking the edges more closely to each other, and hence the dispersion of the segmentation had been lower. This would translate into lower uncertainty of spatial segmentation, and hence higher confidence and reliability of the generated ground truth. Figure 6 shows the distribution of $EV_{max}$ and $EV_{median}$ at different stages of the framework.

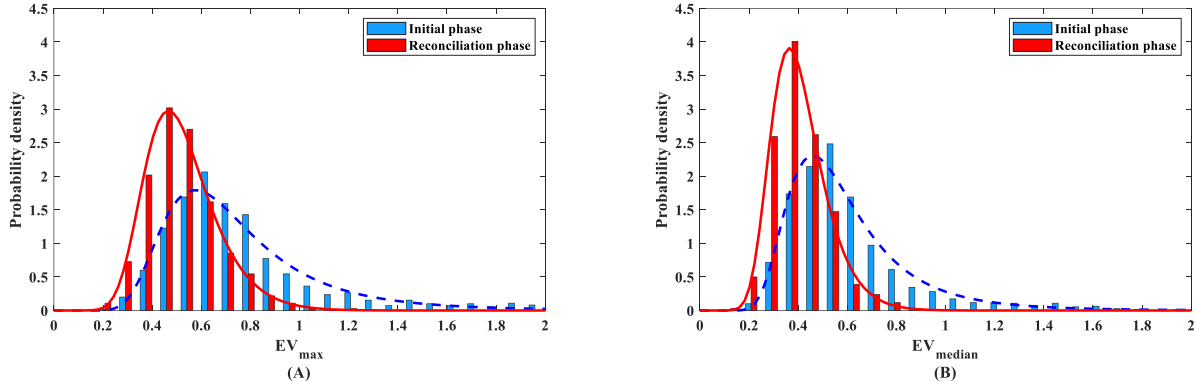

Figure 6. Uncertainty of the ground truth generated at different stages of the framework based of two different criteria. (A) average maximum edge variability ($EV_{max}$), (B) average median edge variability ($EV_{median}$)

Figure 6 indicates that uncertainty of the ground truth decreased as we moved through different stages of the framework. $EV_{max}$ had non-gaussian distributions based on the Shapiro-Wilk test ($p<0.00001$); therefore, the Kruskal-Wallis test formally evaluated significant differences between the uncertainty of different stages of the framework. The independent variable was one of the uncertainty scores ($EV_{max}$ and $EV_{median}$), and the dependent variable was different stages of the framework. There was a significant difference in $EV_{max}$ across the stages of the framework ($\chi^2(1) = 540.8$, $p<0.00001$) with a relatively strong effect size ($\varepsilon^2=0.20$). Post-hoc analysis confirmed significantly lower $EV_{max}$ after reconciliation compared to the initial phase. Similarly, there was a significant difference in $EV_{median}$ across the stages of the framework ($\chi^2(1) = 611.9$, $p<0.00001$) with a relatively strong effect size ($\varepsilon^2=0.23$). Post-hoc analysis confirmed significantly lower $EV_{median}$ after reconciliation compared to the initial phase.

DISCUSSIONS

The aims of this study were (1) to demonstrate that different experts will segment vocal fold edges quite differently and hence to provide quantitative evidence that spatial segmentation is similar to other subjective tasks in our field and requires participation of multiple raters, evaluation of intra-rate reliability, and evaluation of inter-rate reliability. (2) To present a framework for generating reliable spatial segmentation ground truth with sub-pixel resolution. The results of Figures 4(A), 4(B), 5(A), and 5(B) and their formal statistical analysis confirmed that one of our experts was segmenting the vocal fold edges differently from the other experts. This indicates that the ground truth generated from the segmentation of that expert would be very different from the other two experts. Such a finding could not have been determined if multiple raters were not included in a process that included a rigorous evaluation of inter- and intra-rater reliabilities. Stated differently, there is no way to assess the "quality" of the ground truth generated from the manual segmentation of one rater. However, the most interesting finding was that the proposed framework handled significant differences between experts very effectively during the reconciliation phase. Specifically, the relatively strong effect of intra-rater $EV_{max}$ in the initial phase was reduced to a weak effect size after reconciliation that was only marginally significant. We found similar results for inter-rater reliability scores. The moderate effect size of inter-rater $EV_{max}$ in the initial phase was reduced to a negligible effect size after reconciliation.

Results of Figure 6 and its subsequent formal statistical analyses confirmed that generated ground truth became more reliable as we progressed through the stages of the proposed framework. Specifically, $EV_{max}$ and $EV_{median}$ of manual segmentations used to generate the ground truth were significantly lower after the reconciliation phase compared to the initial phase. This means segmented edges were significantly closer to each other after the reconciliation phase, indicating significantly higher confidence and reliability for the generated ground truth.

CONCLUSION

The outcome of the manual segmentation is the gold standard and is used to evaluate the performance of the automated spatial segmentation method, and/or is used to create and tune the automated method. This study quantified possible differences in manual segmentation of vocal fold edges among three different experts and showed they performed the task very differently. We presented a framework that allowed multiple experts to participate in the manual segmentation of laryngeal images in an iterative and multi-stage procedure. The presented framework incorporated the evaluation of inter- and intra-rater reliability and governed generation of very reliable ground truth with sub-pixel resolution. Results of statistical analyses confirmed the effectiveness of the presented framework in handling differences among experts where significant differences with moderate effect sizes during the initial phase were reduced to negligible effect sizes after the reconciliation phase. Additionally, segmented edges were significantly closer to each other after the reconciliation phase, indicating significantly higher confidence in the ground truth that was generated after the reconciliation phase.

ACKNOWLEDGMENTS

This research was funded by the National Institutes of Health - National Institute on Deafness and Other Communication Disorders (R01 DC017923, R01 DC019402DD).

REFERENCES

[1] R. R. Patel *et al.*, "Recommended Protocols for Instrumental Assessment of Voice: American Speech-Language-Hearing Association Expert Panel to Develop a Protocol for Instrumental Assessment of Vocal Function," *Am J Speech Lang Pathol*, vol. 27, no. 3, pp. 887–905, 2018, doi: 10.1044/2018_ajslp-17-0009.

[2] D. D. Deliyski, M. E. G. Powell, S. R. C. Zacharias, T. T. Gerlach, and A. De Alarcon, "Experimental investigation on minimum frame rate requirements of high-speed videoendoscopy for clinical voice assessment," *Biomed Signal Process Control*, vol. 17, pp. 21–28, 2015, doi: 10.1016/j.bspc.2014.11.007.

[3] D. D. Deliyski, P. P. Petrushev, H. S. Bonilha, T. T. Gerlach, B. Martin-Harris, and R. E. Hillman, "Clinical implementation of laryngeal high-speed videoendoscopy: Challenges and evolution," *Folia Phoniatrica et Logopaedica*, vol. 60, no. 1, pp. 33–44, 2008.

[4] H. Ghasemzadeh, D. D. Deliyski, R. E. Hillman, and D. D. Mehta, "Method for horizontal calibration of laser-projection transnasal fiberoptic high-speed videoendoscopy," *Applied Sciences*, vol. 11, no. 2, p. 822, 2021, doi: 10.3390/app11020822.

[5] H. Ghasemzadeh, D. D. Deliyski, R. E. Hillman, and D. D. Mehta, "Framework for indirect spatial calibration of the horizontal plane of endoscopic laryngeal images," *Journal of Voice (In press)*, 2021.

[6] M. Semmler, S. Kniesburges, V. Birk, A. Ziethe, R. Patel, and M. Döllinger, "3D reconstruction of human laryngeal dynamics based on endoscopic high-speed recordings," *IEEE Trans Med Imaging*, vol. 35, no. 7, pp. 1615–1624, 2016.

[7] H. Ghasemzadeh, D. Deliyski, D. Ford, J. B. Kobler, R. E. Hillman, and D. D. Mehta, "Method for Vertical Calibration of Laser-Projection Transnasal Fiberoptic High-Speed Videoendoscopy," *Journal of Voice*, vol. 34, no. 6, pp. 847–861, 2020.

[8] D. Deliyski and P. Petrushev, "Methods for objective assessment of high-speed videoendoscopy," *Proc. Advances in Quantitative Laryngology (AQL)*, pp. 1–16, 2003.

[9] D. Deliyski, "Laryngeal high-speed videoendoscopy," in *Laryngeal evaluation: Indirect laryngoscopy to high-speed digital imaging*, K. Kendall and R. Leonard, Eds., Thieme Medical, New York, NY, 2010, pp. 245–270.

[10] H. Ghasemzadeh, *Quantitative Methods for Calibrated Spatial Measurements of Laryngeal Phonatory Mechanisms*. Michigan State University, 2020.

[11]  G. Andrade-Miranda, Y. Stylianou, D. D. Deliyski, J. I. Godino-Llorente, and N. Henrich Bernardoni, "Laryngeal image processing of vocal folds motion," *Applied Sciences*, vol. 10, no. 5, p. 1556, 2020.

[12]  H. Ghasemzadeh and D. Deliyski, "Non-linear image distortions in flexible fiberoptic endoscopes and their effects on calibrated horizontal measurements," *Journal of Voice*, p. [Epub ahead of print], 2020.

[13]  T.-Y. Hsiao, C.-L. Wang, C.-N. Chen, F.-J. Hsieh, and Y.-W. Shau, "Noninvasive assessment of laryngeal phonation function using color Doppler ultrasound imaging," *Ultrasound Med Biol*, vol. 27, no. 8, pp. 1035–1040, 2001.

[14]  R. Patel, K. D. Donohue, H. Unnikrishnan, and R. J. Kryscio, "Kinematic measurements of the vocal-fold displacement waveform in typical children and adult populations: quantification of high-speed endoscopic videos," *Journal of Speech, Language, and Hearing Research*, vol. 58, no. 2, pp. 227–240, 2015.

[15]  V. S. McKenna, M. E. Diaz-Cadiz, A. C. Shembel, N. M. Enos, and C. E. Stepp, "The relationship between physiological mechanisms and the self-perception of vocal effort," *Journal of Speech, Language, and Hearing Research*, vol. 62, no. 4, pp. 815–834, 2019.

[16]  V. S. McKenna, E. S. Heller Murray, Y.-A. S. Lien, and C. E. Stepp, "The relationship between relative fundamental frequency and a kinematic estimate of laryngeal stiffness in healthy adults," *Journal of Speech, Language, and Hearing Research*, vol. 59, no. 6, pp. 1283–1294, 2016.

[17]  T. Iwahashi, M. Ogawa, K. Hosokawa, C. Kato, and H. Inohara, "A detailed motion analysis of the angular velocity between the vocal folds during throat clearing using high-speed digital imaging," *Journal of Voice*, vol. 30, no. 6, pp. 770.e1-770.e8, 2016.

[18]  H. Ghasemzadeh, R. E. Hillman, and D. D. Mehta, "Toward Generalizable Machine Learning Models in Speech, Language, and Hearing Sciences: Estimating Sample Size and Reducing Overfitting," *Journal of Speech, Language, and Hearing Research*, vol. 67, no. 3, pp. 753–781, 2024.

[19]  D. D. Mehta, D. D. Deliyski, T. F. Quatieri, and R. E. Hillman, "Automated measurement of vocal fold vibratory asymmetry from high-speed videoendoscopy recordings," *Journal of Speech, Language, and Hearing Research*, vol. 54, no. 1, pp. 47–54, 2011.

[20]  Y. Yan, X. Chen, and D. Bless, "Automatic tracing of vocal-fold motion from high-speed digital images," *IEEE Trans Biomed Eng*, vol. 53, no. 7, pp. 1394–1400, 2006.

[21]  T. Wittenberg, M. Moser, M. Tigges, and U. Eysholdt, "Recording, processing, and analysis of digital high-speed sequences in glottography," *Mach Vis Appl*, vol. 8, pp. 399–404, 1995.

[22]  J. Lohscheller, H. Toy, F. Rosanowski, U. Eysholdt, and M. Döllinger, "Clinically evaluated procedure for the reconstruction of vocal fold vibrations from endoscopic digital high-speed videos," *Med Image Anal*, vol. 11, no. 4, pp. 400–413, 2007.

[23]  A. M. Yousef, D. D. Deliyski, S. R. C. Zacharias, A. de Alarcon, R. F. Orlikoff, and M. Naghibolhosseini, "Spatial Segmentation for Laryngeal High-Speed Videoendoscopy in Connected Speech," *Journal of Voice*, 2020.

[24]  P. Gómez *et al.*, "BaGLS, a multihospital Benchmark for automatic Glottis Segmentation," *Sci Data*, vol. 7, no. 1, pp. 1–12, 2020.

[25]  M. V. A. Rao, R. Krishnamurthy, P. Gopikishore, V. Priyadharshini, and P. K. Ghosh, "Automatic Glottis Localization and Segmentation in Stroboscopic Videos Using Deep Neural Network.," in *INTERSPEECH*, 2018, pp. 3007–3011.

[26]  D. Britton *et al.*, "Endoscopic assessment of vocal fold movements during cough," *Annals of Otology, Rhinology & Laryngology*, vol. 121, no. 1, pp. 21–27, 2012.

[27]  H. J. Moukalled, D. D. Deliyski, R. R. Schwarz, and S. Wang, "Segmentation of Laryngeal High-Speed Videoendoscopy in Temporal Domain Using Paired Active Contours," *Sixth International Workshop on*

*Models and Analysis of Vocal Emissions for Biomedical Applications, MAVEBA*, vol. 9, no. d, pp. 137–140, 2009.

[28] A. M. Yousef, D. D. Deliyski, S. R. C. Zacharias, A. de Alarcon, R. F. Orlikoff, and M. Naghibolhosseini, "A deep learning approach for quantifying vocal fold dynamics during connected speech using laryngeal high-speed videoendoscopy," *Journal of Speech, Language, and Hearing Research*, vol. 65, no. 6, pp. 2098–2113, 2022.

[29] G. Andrade-Miranda, J. I. Godino-Llorente, L. Moro-Velázquez, and J. A. Gómez-Garcia, "An automatic method to detect and track the glottal gap from high speed videoendoscopic images," *Biomed Eng Online*, vol. 14, pp. 1–29, 2015.

[30] M. V. A. Rao, R. Krishnamurthy, P. Gopikishore, V. Priyadharshini, and P. K. Ghosh, "Automatic Glottis Localization and Segmentation in Stroboscopic Videos Using Deep Neural Network.," in *INTERSPEECH*, 2018, pp. 3007–3011.

[31] M. Motie-Shirazi, M. Zañartu, S. D. Peterson, and B. D. Erath, "Vocal fold dynamics in a synthetic self-oscillating model: Contact pressure and dissipated-energy dose," *J Acoust Soc Am*, vol. 150, no. 1, pp. 478–489, 2021.

[32] M. E. Powell *et al.*, "Efficacy of Videostroboscopy and High-Speed Videoendoscopy to Obtain Functional Outcomes From Perioperative Ratings in Patients With Vocal Fold Mass Lesions," *Journal of Voice*, vol. 34, no. 5, pp. 769–782, 2020, doi: 10.1016/j.jvoice.2019.03.012.

[33] D. Rao, P. K, and R. Singh, "Automated segmentation of the larynx on computed tomography images: a review," *Biomed Eng Lett*, vol. 12, no. 2, pp. 175–183, 2022.

[34] P. Jaccard, "The distribution of the flora in the alpine zone," *New phytologist*, vol. 11, no. 2, pp. 37–50, 1912.

[35] J. Jcgm and others, "Evaluation of measurement data—Guide to the expression of uncertainty in measurement," *Int. Organ. Stand. Geneva ISBN*, vol. 50, p. 134, 2008.